\documentclass[aps,prc,twocolumn,superscriptaddress]{revtex4-1}
\usepackage{CJK}
\usepackage{graphicx}
\usepackage{hyperref}
\usepackage{amssymb}
\usepackage{amsmath}
\usepackage[normalem]{ulem}
\usepackage{url}
\usepackage{color}

\begin{document}
\begin{CJK*} {UTF8}{} 

\title{Constraints on the symmetry energy and its associated parameters from nuclei to neutron stars}

\author{Yingxun Zhang} 
\email{zhyx@ciae.ac.cn}
\affiliation{China Institute of Atomic Energy, Beijing 102413, P.R. China}
\affiliation{Guangxi Key Laboratory Breeding Base of Nuclear Physics and Technology, Guilin 541004, China}





\author{Min Liu} 
\affiliation{Guangxi Normal University, Guilin, 541004, P.R.China}
\author{Cheng-Jun Xia}
\affiliation{School of Information Science and Engineering, Zhejiang University Ningbo Institute of Technology, Ningbo 315100, China}
\author{Zhuxia Li} 
\affiliation{China Institute of Atomic Energy, Beijing 102413, P.R. China}
\author{Subrata Kumar Biswal} 
\affiliation{Department of Astronomy, Xiamen University, Xiamen 361005, P.R. China}


\date{\today}

\begin{abstract}
The symmetry energy obtained with the effective Skyrme energy density functional is related to the values of isoscalar effective mass and isovector effective mass, which is also indirectly related to the incompressibility of symmetric nuclear matter. In this work, we analyze the values of symmetry energy and its related nuclear matter parameters in five-dimensional parameter space by describing the heavy ion collision data, such as isospin diffusion data at 35 MeV/u and 50 MeV/u, neutron skin of $^{208}$Pb, and tidal deformability and maximum mass of neutron star. We obtain the parameter sets which can describe the isospin diffusion, neutron skin, tidal deformability and maximum mass of neutron star, and give the incompressibility $K_0$=250.23$\pm$20.16 MeV, symmetry energy coefficient $S_0$=31.35$\pm$2.08 MeV, the slope of symmetry energy $L$=59.57$\pm$10.06 MeV, isoscalar effective mass $m_s^*/m$=0.75$\pm$0.05 and quantity related to effective mass splitting $f_I$=0.005$\pm$0.170. At two times normal density, the symmetry energy we obtained is in 35-55 MeV. To reduce the large uncertainties of $f_I$, more critical works in heavy ion collisions at different beam energies are needed.
\end{abstract}

\pacs{21.60.Jz, 21.65.Ef, 24.10.Lx, 25.70.-z}

\maketitle
\end{CJK*}

\section{Introduction}
The symmetry energy describes the energy related to the excess neutrons or protons in a nuclear system, which tends to reach the isospin symmetry with $N=Z$ and is of fundamental importance in our understanding of nature's asymmetric objects including neutron stars~\cite{Lattimer00,Abbott17,Abbott18,Fattoyev18,Annala18,Abbott19,Malik19,NBZhang18,NBZhang19,WJXie19,CYTsang19,MBTsang19,HTong19,YZhou19APJ,YZhou19PRD} as well as heavy nuclei with very different number of neutrons and protons~\cite{BALi97,Tsang04,TXLiu07,Sun10,Baran05,BALi08,Lynch09,QHWu15}. However, theoretical studies show that the density dependence of symmetry energy is the most uncertain part in the isospin asymmetric nuclear equation of state (EOS) ~\cite{Brown00,BALi08}. There have been lots of effort to constrain the density dependence of symmetry energy by using neutron skin~\cite{Brown00,Typel01,Centelles09,LWChen10,ZZhang13}, giant dipole resonance~\cite{Trippa08}, electric dipole polarizability~\cite{Roca13,ZZhang14}, mass-radius relation and tidal deformability of neutron stars~\cite{Abbott17,Abbott18,Fattoyev18,Annala18,Abbott19,Malik19,NBZhang18,NBZhang19,WJXie19,CYTsang19,MBTsang19,HTong19,YZhou19APJ,YZhou19PRD}, and heavy ion collisions (HICs)~\cite{LWChen05,Tsang09,Russotto16}, and consensus on the symmetry energy coefficients $S_0=S(\rho_0)$, and the slope of symmetry energy $L=3\rho_0 \frac{\partial S(\rho)}{\partial \rho}|_{\rho_0}$ have been obtained but with different uncertainties ~\cite{Lattimer14}. Here, $S(\rho)$ is the density dependence of the symmetry energy and its Taylor expansion around normal density is
\begin{eqnarray}
\label{srhotaylor}
S(\rho)&=&S_0+L(\frac{\rho-\rho_0}{3\rho_0})+\frac{K_{sym}}{2}(\frac{\rho-\rho_0}{3\rho_0})^2\\\nonumber
&&+\frac{1}{6}Q_{sym}(\frac{\rho-\rho_0}{3\rho_0})^3\cdots,
\end{eqnarray}
$K_{sym}$ and $Q_{sym}$ are the curvature and skewness parameters of $S(\rho)$. There are also many efforts to constrain the $K_{sym}$ and $Q_{sym}$ from neutron skin and neutron star~\cite{WJXie19,Margueron18,Margueron19,NBZhang19}.

However, Margueron $et al.$'s calculations show that the simple Taylor expansion of the EOS cannot be used to reproduce the EOS well at the whole density region as well as for the symmetry energy, and they proposed a meta-EOS model to describe it~\cite{Margueron18,Margueron19,NBZhang19}. Another method to well describe the Skyrme EOS and symmetry energy is to use the nuclear matter parameters, such as $\rho_0$, $E_0$, $K_0$, $S_0$, $L$, $m_s^*$, $m_v^*$, with two additional coefficients $g_{sur}$ and $g_{sur,iso}$~\cite{Agrawal05,LWChen09,Zhang15}. Here, $\rho_0$ is the normal density, $K_0=9\rho_0 \frac{\partial^2 \epsilon/\rho}{\partial \rho^2 } |_{\rho_0}$ is the incompressibility of symmetric nuclear matter, $m_s^*/m=(1+\frac{2m}{\hbar^2} \frac{\partial }{\partial \tau} \frac{E}{A})|_{\rho_0 }$ is the isoscalar effective mass, $m_v^*=\frac{1}{1+\kappa}$ is the isovector effective mass where $\kappa$ is the enhancement of a factor of the Thomas-Reich-Kuhn sum rule. $g_{sur}$, $g_{sur,iso}$ are the coefficients related to density gradient terms. A lot of theoretical works have evidenced that all of them are related to the symmetry energy. For example, in the Skyrme-Hartree-Fock approaches, the density dependence of symmetry energy is written as,
\begin{eqnarray}
\label{srhoshf}
S(\rho)&=&\frac{1}{3}\frac{\hbar^2}{2m}(\frac{3\pi^2}{2}\rho)^{2/3}\\\nonumber
&&+(A_{sym}u+B_{sym}u^\eta+C_{sym}(m_s^*,m_v^*)u^{5/3}),\\\nonumber
\end{eqnarray}
where $u$ is the reduced density, i.e., $\rho/\rho_0$. In the right-hand side of Eq.~(\ref{srhoshf}), the first term comes from the kinetic energy contribution, the second and third terms are from the two-body and effective three-body interactions, the fourth term is from the momentum dependent interaction and is related to $m_s^*$ and $m_v^*$. A recent theoretical study by Mondal and Agrawal $et al$. also provide evidence that the $S(\rho)$ depends on the effective mass~\cite{Mondal17}. Thus, one can expect that the constraint of $S(\rho)$ with less biased uncertainty should depend on the values of $\rho_0$, $E_0$, $K_0$, $S_0$, $L$, $m_s^*$, $m_v^*$ rather than only on the uncertainties of $S_0$ and $L$.

In this work, we adopt the five nuclear matter parameters $K_0$, $S_0$, $L$, $m_s^*$, $f_I$ as inputs at given the values of $\rho_0$, $E_0$, $g_{sur}$, and $g_{sur,iso}$, because the nuclear matter parameters, such as $K_0$, $S_0$, $L$, $m_s^*$, $m_v^*$, still have certain uncertainties~\cite{Dutra12}. In the transport model calculations, we replace $m^*_v$ by $f_I$, which is defined as $f_I=\frac{1}{2\delta}(\frac{m}{m_n^*}-\frac{m}{m_p^*})=\frac{m}{m_s^*}-\frac{m}{m_v^*}$, since the $f_I$ can be analytically incorporated into the transport model and its sign reflects the $m^*_n>m_p^*$ or $m^*_n<m_p^*$.
We finally give the range of nuclear matter parameters $K_0$, $S_0$, $L$, $m_s^*$, $f_I$, which are estimated based on the description of isospin diffusion data, the neutron skin of $^{208}$Pb, and tidal deformability and maximum mass of the neutron star.

\section{Theoretical Models}
\subsection{ImQMD model}
The transport model used in this work is the ImQMD-Sky~\cite{Zhang14,Zhang15}. In the model, the nucleonic potential energy density without the spin-orbit term is $u_{loc}+u_{md}$, and
\begin{eqnarray}
\label{eq:edfimqmd}
u_{loc}=&&\frac{\alpha}{2}\frac{\rho^2}{\rho_0} +\frac{\beta}{\eta+1}\frac{\rho^{\eta+1}}{\rho_0^\eta}+\frac{g_{sur}}{2\rho_0 }(\nabla \rho)^2+\\\nonumber
&&\frac{g_{sur,iso}}{\rho_0}[\nabla(\rho_n-\rho_p)]^2+A_{sym}\frac{\rho^2}{\rho_0}\delta^2+B_{sym}\frac{\rho^{\eta+1}}{\rho_0^\eta}\delta^2
\end{eqnarray}
and Skyrme-type momentum dependent energy density functional $u_{md}$ is written based on its interaction form $\delta (\mathbf r_1-\mathbf r_2 ) (\mathbf p_1-\mathbf p_2 )^2$~\cite{Skyrme56, Vauthe72,Zhang15} as,
\begin{eqnarray}
\label{eq:mdimqmd}
u_{md}=&&C_0\sum_{ij}\int d^3pd^3p' f_i(\mathbf r,\mathbf p)f_j(\mathbf r,\mathbf p')(\mathbf p-\mathbf p')^2+\\\nonumber
&&D_0\sum_{ij\in n}\int d^3pd^3p'f_i(\mathbf r,\mathbf p) f_j(\mathbf r,\mathbf p')(\mathbf p-\mathbf p')^2 +\\\nonumber
&&D_0\sum_{ij\in p}\int d^3p d^3p' f_i(\mathbf r,\mathbf p)f_j(\mathbf r,\mathbf p')(\mathbf p-\mathbf p')^2.
\end{eqnarray}
The connection between nine parameters $\alpha$, $\beta$, $\eta$, $A_{sym}$, $B_{sym}$, $C_0$, $D_0$, $g_{sur}$, $g_{sur,iso}$ used in ImQMD-Sky and the nine nuclear matter parameters, $\rho_0$, $E_0$, $K_0$, $S_0$, $L$, $m_s^*$, $ m_v^*$, $g_{sur}$, $g_{sur,iso}$, are given by the following analytical relationship,
\begin{eqnarray}
&& g_{\rho\tau}=\frac{3}{5}(\frac{m_0}{m_s^*}-1)\epsilon_F^0, \\\nonumber &&\eta=(K_0+\frac{6}{5}\epsilon_F^0-10g_{\rho\tau})/(\frac{9}{5}\epsilon_F^0-6g_{\rho\tau}-9E_0)\\\nonumber
&&\beta=\frac{(\frac{1}{5}\epsilon_F^0-\frac{2}{3} g_{\rho\tau}-E_0 )(\eta+1)}{\eta-1}, \alpha=E_0-\epsilon_F^0-\frac{8}{3} g_{\rho\tau}-\beta,\\\nonumber
&& C_0=\frac{1}{16\hbar^2}\Theta_v, D_0=\frac{1}{16\hbar^2}(\Theta_s-2\Theta_v),\\\nonumber
&&C_{sym}=-\frac{1}{24}(\frac{3\pi^2}{2})^{2/3} (3\Theta_v-2\Theta_s )\rho_0^{5/3},\\\nonumber
&&B_{sym}=\frac{3S_0-L-\frac{1}{3}\epsilon_F^0+2C_{sym} (m_s^*,m_v^* )}{-3\sigma}\\\nonumber
&&A_{sym}=S_0-\frac{1}{3}\epsilon_F^0-B_{sym}-C_{sym} (m_s^*,m_v^*)
\label{eq:skyqmd}
\end{eqnarray}
where $\Theta_s=(\frac{m_0}{m_s^*}-1) \frac{8\hbar^2}{m_0\rho_0}$, $\Theta_v=(\frac{m_0}{m_v^*}-1)\frac{4\hbar^2}{m_0\rho_0}$, and $\eta=\sigma+1$. A similar relation has been discussed in Refs.~\cite{Agrawal05,LWChen09}. The approach used in this work is that we set the nine nuclear matter parameters $\rho_0$, $E_0$, $K_0$, $S_0$, $L$, $m_s^*$, $m_v^*$, $g_{sur}$, $g_{sur,iso}$ as the input of the ImQMD-Sky code. The coefficients of the density gradient terms are set as $g_{sur} = 24.5$ MeVfm$^2$ and $g_{sur;iso} = -4.99$ MeVfm$^2$, and varying of $g_{sur}$ and $g_{sur,iso}$ in a reasonable region for different Skyrme interactions has negligible effects on the calculated experimental observables in intermediate energy heavy ion collisions. The nucleon-nucleon collision and Pauli-blocking part used in this work are treated as the same as those in Refs.~\cite{Zhang05,Zhang06,Zhang07}, and we do not vary its strength or form in this study since previous calculations have shown it does not strongly influence the isospin sensitive observables we studied~\cite{Zhang12}.

\subsection{Density variational method}
The approach we used to calculate the neutron skin is the restricted density variational method (RDV), which is the same as in Ref.~\cite{MLiu06}, where the semiclassical expressions of the Skyrme energy density functional are applied to study the ground state energies, the neutron proton density distributions, and the neutron skin thickness of a series of nuclei. The binding energy of a nucleus is expressed as the integral of energy density functional, i.e.,
\begin{equation}
\label{eq:edf}
E=\int \mathcal{H} dr=\int \frac{\hbar^2}{2m}[\tau_n(\mathbf{r})+\tau_p(\mathbf{r})]+\mathcal{H}_{sky}+\mathcal{H}_{coul} dr.
\end{equation}
The $\mathcal{H}_{sky}$ is nucleonic density functional, which has the same form as we used in the ImQMD model, but with the spin-orbit interaction form and $W_0$=130 MeVfm$^{5}$. The kinetic energy density in the RDV method is given by
\begin{eqnarray}
\label{eq:tau}
\tau_i(\mathbf{r})&=&\frac{3}{5}(3\pi^2)^{2/3}\rho_i^{5/3}+\frac{1}{36}\frac{(\nabla\rho_i)^2}{\rho_i}+\frac{1}{3}\triangle\rho_i\\\nonumber
&&+\frac{1}{6}\frac{\nabla\rho_i+\nabla f_i+\rho_i\triangle f_i}{f_i}-\frac{1}{12}\rho_i(\frac{\nabla f_i}{f_i})^2\\\nonumber
&&+\frac{1}{2}\rho_i(\frac{2m}{\hbar^2}\frac{W_0}{2}\frac{\nabla(\rho+\rho_i)}{f_i})^2,
\end{eqnarray}
where the extended Thomas-Fermi (ETF) approach including all terms up to second order (ETF2) and fourth order (ETF4) as in Ref.~\cite{Brack85}. $\rho_i$ denotes the proton and neutron density of nucleus, and $\rho=\rho_n+\rho_p$. $W_0$ is the strength of the spin-orbit interaction; the parameter $f_i(\mathbf{r})$ is the same as in Ref.~\cite{MLiu06}. The Coulomb energy density is written as the sum of the direct and exchange terms.
In the calculations, we take the density distribution as a spherical symmetric Fermi function:
\begin{equation}
\rho_i=\rho_{0i}[1+\exp(\frac{r-R_{0i}}{a_i})], i={n, p}.
\end{equation}
Here, $R_{0p}$, $a_p$, $R_{0n}$, and $a_n$ are the radius and diffuseness of proton and neutron density distributions. By minimizing the total energy of the system given by Eq.~(\ref{eq:edf}), the neutron and proton densities can be obtained and thus the neutron skin. The values of the neutron skin of $^{208}$Pb we obtained are consistent with the results obtained with Skyrme Hartree-Fock calculations~\cite{MLiu06,Tsang19p} after considering the fourth order in extended Thomas-Fermi approach.

\subsection{Tolman-Oppenheimer-Volkov equation}
The structure of neutron stars is obtained by solving the Tolman-Oppenheimer-Volkov equation~\cite{Lipunov92}
\begin{eqnarray}
&&\frac{\mbox{d}P}{\mbox{d}r} = -\frac{G M \epsilon}{r^2}   \frac{(1+P/\epsilon)(1+4\pi r^3 P/M)} {1-2G M/r},  \label{eq:TOV}\\
&&\frac{\mbox{d}M(r)}{\mbox{d}r} = 4\pi \epsilon r^2, \label{eq:m_star}
\end{eqnarray}
while the tidal deformability~\cite{Damour09, Hinderer10,Postnikov10} is estimated with
\begin{equation}
\Lambda = \frac{2 k_2}{3}\left( \frac{R}{G M} \right)^5. \label{eq:td}
\end{equation}
Here, the gravity constant is taken as $G=6.707\times 10^{-45}\ \mathrm{MeV}^{-2}$, $r$ is the distance from the core of the star, $\epsilon=\epsilon(r)$ is the energy density or mass density, $P=P(r)$ is the pressure, and $M=M(r)$ is the mass within the radius $r$. $k_2$ is the second Love number and is obtained from the response of the induced quadrupole moment ${\cal Q}_{ij}$ in a static external quadrupolar tidal field ${\cal E}_{ij}$ with ${\cal Q}_{ij} = -k_2 \frac{2R^5}{3G}{\cal E}_{ij}$~\cite{Damour09, Hinderer10, Postnikov10}.

Based on the Skyrme parameters listed in Table~\ref{tab:22skyrmes}, the energy density of nuclear matter is obtained with
\begin{eqnarray}
  \epsilon_\mathrm{NM}/\rho &=& m + \frac{3\hbar^2}{10 M} \left( \frac{3\pi^2}{2} \rho \right)^{2/3} H_{5/3}\\\nonumber
                      &&     + \frac{t_0}{8} \rho \left[ 2 x_0 + 4- \left( 2 x_0+1 \right) H_2 \right] \nonumber\\
                     &&{}  + \frac{t_3}{48} \rho^{\sigma+1} \left[ 2 x_3 + 4 - \left( 2 x_3+1 \right) H_2 \right]\nonumber\\
 && + \frac{3 \rho}{40} \left( \frac{3\pi^2}{2} \rho\right)^{2/3} \left( a H_{5/3} + b H_{8/3}  \right),\nonumber
 \label{eq:E_NM}
\end{eqnarray}
where $a = t_1 \left( x_1+2 \right) + t_2 \left( x_2+2 \right)$, $b = \frac{1}{2}\left[t_2\left(2 x_2+1\right)-t_1 \left(2 x_1+1\right)\right]$, and $H_n = 2^{n-1} \left[ y^n + \left( 1-y \right)^n \right]$ with $y=\rho_p/\rho$ being the proton fraction. Meanwhile, the energy density of electrons and muons are given by
\begin{eqnarray}
\epsilon_{e,\mu} &=& \int_0^{\nu_{e,\mu}} \frac{p^2}{\pi^2} \sqrt{p^2+m_{e,\mu}^2}\mbox{d}p\\\nonumber
 &=&  \frac{m_{e,\mu}^4}{8\pi^{2}} \left[x_{e,\mu}(2x_{e,\mu}^2+1)\sqrt{x_{e,\mu}^2+1}-\mathrm{arcsh}(x_{e,\mu}) \right]. \label{eq:E_l}
\end{eqnarray}
Here, $x_{e,\mu}\equiv \nu_{e,\mu}/m_{e,\mu}$ with $\nu_{e,\mu}$ being the Fermi momentum of leptons, which predicts their number densities $\rho_{e,\mu} = \nu_{e,\mu}^3/3\pi^2$. The total energy density of neutron star matter is obtained with $\epsilon=\epsilon_\mathrm{NM}+\epsilon_{e}+\epsilon_{\mu}$. Then the pressure is determined by $P = \sum_i \mu_i \rho_i  - \epsilon$ with the chemical potential $\mu_i = \frac{\partial \epsilon}{\partial \rho_i}$. The equation of state (EoS) of neutron star matter in the density range $0.5\rho_0<\rho<3\rho_0$ is obtained by simultaneously fulfilling the $\beta$-stability condition $\mu_n-\mu_p=\mu_e=\mu_\mu$ and local charge neutrality condition $\rho_{e}+\rho_{\mu}=\rho_p$.

At subsaturation densities, the pasta phases of nuclear matter emerge, we thus adopt the EoSs presented in Refs.~\cite{Feynman49, Baym71, Negele73} at $\rho<0.08\ \mathrm{fm}^{-3}$. For the density region above $3\rho_0$, we adopt a polytropic EoS~\cite{Lattimer16,Fattoyev13,CYTsang19}, where the pressure is given by $P =\kappa \rho^{\gamma}$. At given $\gamma$, the parameter $\kappa$ and energy density is fixed according to the continuity condition of pressure and baryon chemical potential at $\rho=3\rho_0$. In this work, we adopt a maximum value with $\gamma=2.9$ so that the velocity of sound does not exceed the speed of light.

\section{Results and Discussions}
\subsection{Uncertainties of $K_0$, $S_0$, $L$, $m_s^*/m$, and $f_I$ in effective Skyrme interactions}
\begin{figure}[htbp]
\centering
\includegraphics[angle=0,scale=0.35]{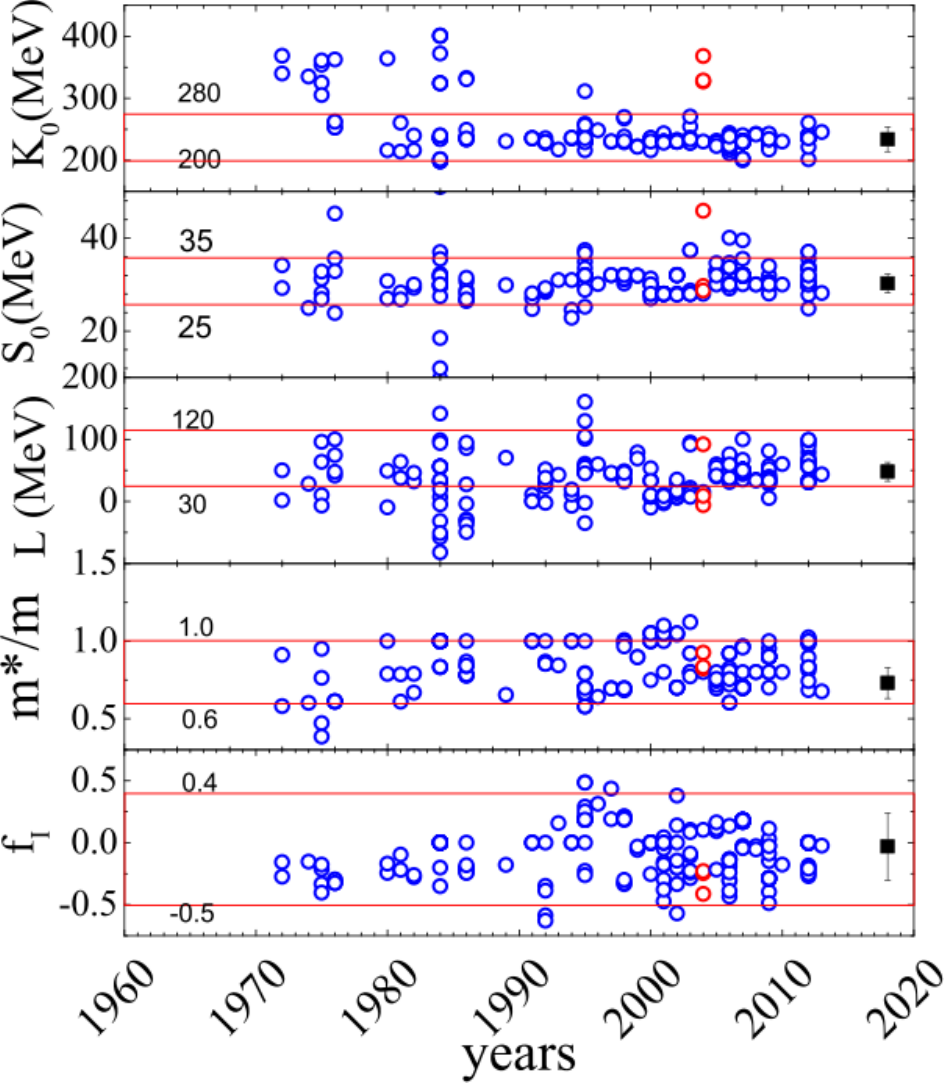}
\setlength{\abovecaptionskip}{0pt}
\caption{\label{sky-prior}(Color online) Extracted values of nuclear matter parameters, $K_0$, $S_0$, $L$, $m_s^*/m$, and $f_I$ as a function of published year. The values are obtained from the compiled Skyrme parameter sets by Dutra $et al$.~\cite{Dutra12}. The black points are the results obtained in~\cite{Margueron18}.
}\label{fig:uncer-5dp}
\setlength{\belowcaptionskip}{0pt}
\end{figure}

Figure~\ref{sky-prior} shows the values of nuclear matter parameters, $K_0$, $S_0$, $L$, $m_s^*/m$, and $f_I$
calculated from 224 effective Skyrme interactions published from the years 1970-2015~\cite{Dutra12}. The nuclear matter incompressibility from Skyrme parameter sets converges to the region of 200-280 MeV after the year $\sim$1990, except for the parameter from the original quark meson coupling (QMC) method~\cite{Guichon04} (red circles in upper panels of Fig.~\ref{sky-prior}) which were readjusted in 2006. For other nuclear matter parameters, such as $S_0$, $L$, $m^*_s$, and $f_I$, most of their values fall into the regions of $S_0=25-35$ MeV, $L=30-120$ MeV, $m^*_s/m=0.6-1.0$, $f_I=-0.5-0.4$. The very recent results on the estimated nuclear matter parameters~\cite{Margueron18} are shown as black squares in Fig.~\ref{fig:uncer-5dp}.

Furthermore, the correlations between the nuclear matter parameters are also very important for getting insight into the effective interaction. For example, as shown in Eq.~(\ref{srhoshf}), $S(\rho)$ comes from the two- and three-body interactions as well as the momentum dependent interaction which is related to $m_s^*$ and $m_v^*$. It means the coefficients of $S_0$ and $L$ should be related to the values of $m_s^*$ and $m_v^*$. The $m_s^*/m$ is also related to the $K_0$ based on the formula of Skyrme Hartree-Fock (SHF) as pointed out in Ref.~\cite{Chab97},
\begin{equation}
K_0=B+C\sigma+D(1-\frac{3}{2}\sigma)\frac{8\hbar^2}{m\rho_0}(\frac{m}{m_s^*}-1),
\end{equation}
with $B=-9E_0+\frac{3}{5}\epsilon_F$, $C=-9E_0+\frac{9}{5}\epsilon_F$ and $D=\frac{3}{20}\rho_0k_F^2$. If the $E_0$ and $\rho_0$ are well known, the $K_0$ depends on the $m_s^*$ and $\sigma$. Focusing on the correlation between $m_s^*$ and $K_0$, one can say $K_0$ is independent of $m_s^*$ if $\sigma=2/3$, but $K_0$ linearly depends on the inverse of $m^*_s$ if $\sigma\ne \frac{2}{3}$.

Quantitatively, we can use the correlation coefficient to understand the correlation among the nuclear matter parameters from the sample of compiled Skyrme parameter sets. We calculate the correlation factor,
\begin{equation}
r_{XY}=\frac{<(X-<X>)(Y-<Y>)>}{\sqrt{<(X-<X>)^2><(Y-<Y>)^2>}},
\end{equation}
where $X$ and $Y$ are two variables we analyzed, and $<X>$ and $<Y>$ are the average values in the selected sample. The values of $r_{XY}$ close to $\pm$1 mean a positive (negative) linear correlation between $X$ and $Y$, and $r_{XY}$ close to zero signifies an essential lack of correlations. By using the Skyrme parameter sets published after the year 2000~\cite{Skysets}, we obtain the correlation factor as follows: $r_{XY}=0.84$ between $S_0$ and $L$, $r_{XY}=-0.35$ between $L$ and $m_s^*$, and  $r_{XY}=-0.34$ between $m_s^*$ and $f_I$. The strongest correlation we obtained from the sample of Skyrme parameter sets is between $S_0$ and $L$.

\begin{figure}[htbp]
\centering
\includegraphics[angle=0,scale=0.25]{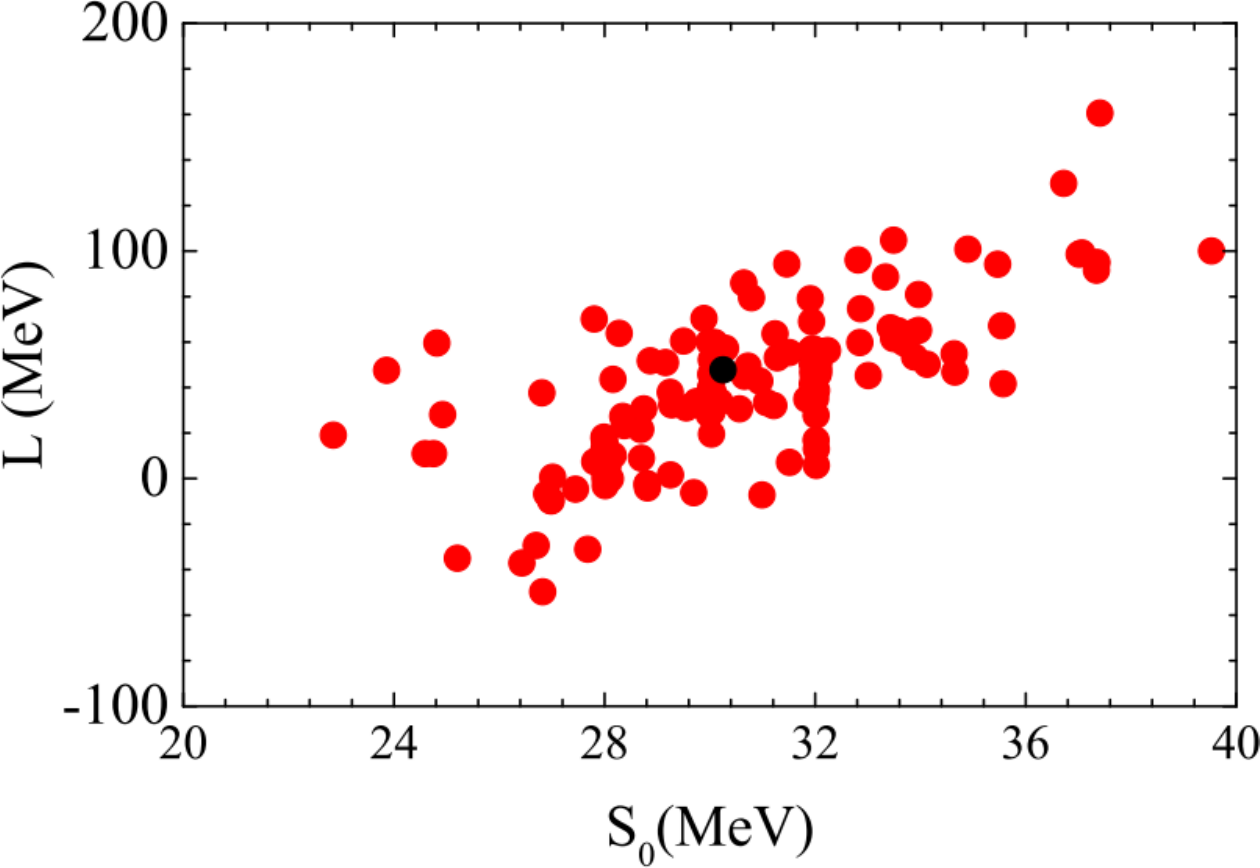}
\setlength{\abovecaptionskip}{0pt}
\caption{\label{nm-corr}(Color online) The correlation between $S_0$ and $L$ obtained from the compilation of Skyrme parameter sets by Dutra $et al$.~\cite{Dutra12} and Ref.~\cite{Margueron18}. Only the results in the region of $S_0$=[20,40]MeV and $L$=[-100,200]MeV are presented. The black points are the results obtained in~\cite{Margueron18}.
}
\setlength{\belowcaptionskip}{0pt}
\end{figure}

By analyzing the slope of $L$ vs $S_0$, one can learn the values of the sensitive density $\rho_s$ which is related to the fitting data that we used to construct the effective Skyrme force~\cite{Lynch18}. 
The high order terms in Eq.~(\ref{srhotaylor}) could weaken this correlation. Figure~\ref{nm-corr} shows the positive correlation between $L$ and $S_{0}$ in the range of $S_0=[20, 40]$MeV and $L=[-100,200]$ MeV obtained from the compilation of Skyrme parameter sets~\cite{Dutra12}. It means the Skyrme parameter sets we used mainly reflect the symmetry energy at subsaturation density. However, one can find that the points spread in a region and they did not fall on the same line. One of the reasons is that the effective Skyrme parameter sets were constructed for best fitting the different observables or nuclei and they may reflect the symmetry energy at different density. For deeply understanding the correlation and narrow the region of $S_0$ and $L$, it is better that one constrains it from many sides, such as from heavy ion collisions, neutron skin, and neutron star, simultaneously. 

\subsection{Isospin diffusion}

The isospin diffusion reflects the changes of isospin asymmetry of the projectile/target-like residue immediately after the peripheral collision and prior to secondary decay for asymmetric reaction system, such as $^{124}$Sn+$^{112}$Sn. It can be measured by the isospin transport ratios $R_i$ in the projectile rapidity region which is defined as
\begin{equation}
R_i=\frac{2X_{ab}-X_{aa}-X_{bb}}{X_{aa}-X_{bb}}.
\label{ridef}
\end{equation}
At least, three reaction systems, two symmetric systems, such as $^{112}$Sn+$^{112}$Sn and $^{124}$Sn+$^{124}$Sn, and one mixing system $^{112}$Sn+$^{124}$Sn or $^{124}$Sn+$^{112}$Sn, should be used. In Eq.(\ref{ridef}), $a=^{124}Sn$, $b=^{112}Sn$, and $X$ is the HIC observable.

\begin{figure}[htbp]
\centering
\includegraphics[angle=0,scale=0.28]{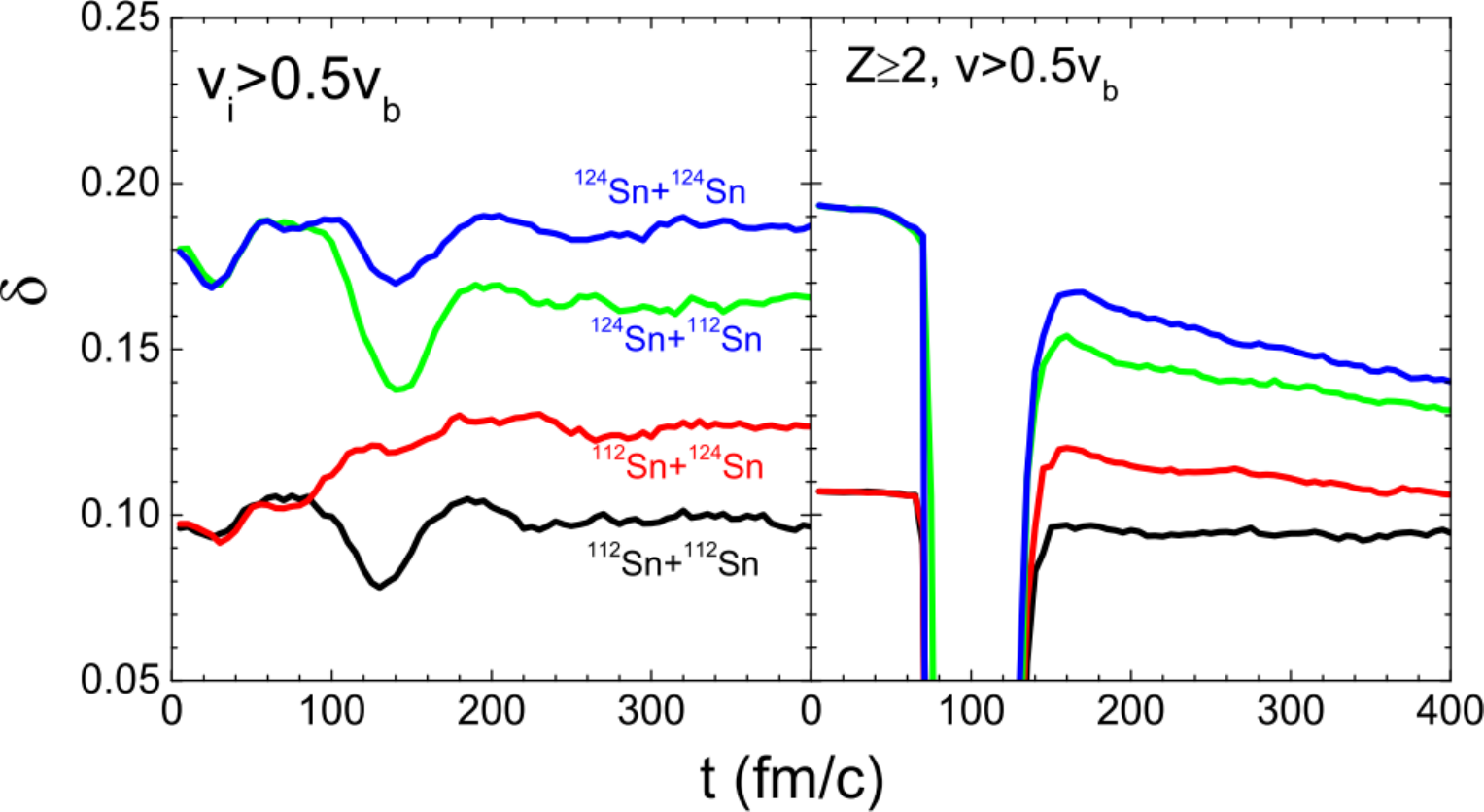}
\setlength{\abovecaptionskip}{0pt}
\caption{\label{deltatime} (Color online) The isospin asymmetry of the `emitting source' as a function of time. Left panel is the emitting source constructed from nucleons and fragments in the projectile rapidity region, and right panel is the emitting source constructed from fragments with $Z\ge2$ in the projectile rapidity region.}
\setlength{\belowcaptionskip}{0pt}
\end{figure}

In transport model simulations, the isospin asymmetry of the ``emitting source''~\cite{Tsang04,LWChen05,Tsang09,Zhang12}, $X=\delta$, is adopted to analyze the isospin diffusion rather than directly use the isoscaling parameter $X=\alpha$~\cite{Tsang04} or $X=ln(Y(^7Li)/Y(^7Be))$~\cite{Sun10, TXLiu07}. The reasons are that: 1) the isospin diffusion reflects the change of isospin asymmetry of the projectile-like residue immediately after the collision and prior to secondary decay, and thus we need the isospin asymmetry of the emitting source at that time; and 2) the definition of `emitting source' should be coalescence invariant, i.e., it can contain all the emitted nucleons or fragments in the late stage, to overcome the deficient of the cluster formation mechanism in the transport model. Based on this concept, the `emitting source' is constructed from the emitted nucleons and fragments with a velocity greater than half of the beam velocity, i.e., $v_i>0.5v_b^{c.m.}$, $i$=fragments, nucleons~\cite{LWChen05,Tsang09,Zhang14,Zhang15}, as the same condition as in experiments. As an example, we illustrate the definition of the `emitting source' we used in the left panel of Fig.~\ref{deltatime}. 
The four lines with different colors refer to the four different reaction systems. It is clear that the isospin asymmetry of the emitting source reaches saturation values after about 200 fm/$c$, which corresponds to the change of isospin asymmetry of the projectile-like residue immediately after the collision and prior to secondary decay. But, if the `emitting source' is constructed from the fragments with $Z\ge 2$, the isospin asymmetry decreases with time due to the nucleons emission and the deficiency of cluster formation and emission of neutrons in transport models. The most important point is that it does not exactly reflect the isospin diffusion as we discussed.


\begin{figure}[htbp]
\centering
\includegraphics[angle=0,scale=0.32]{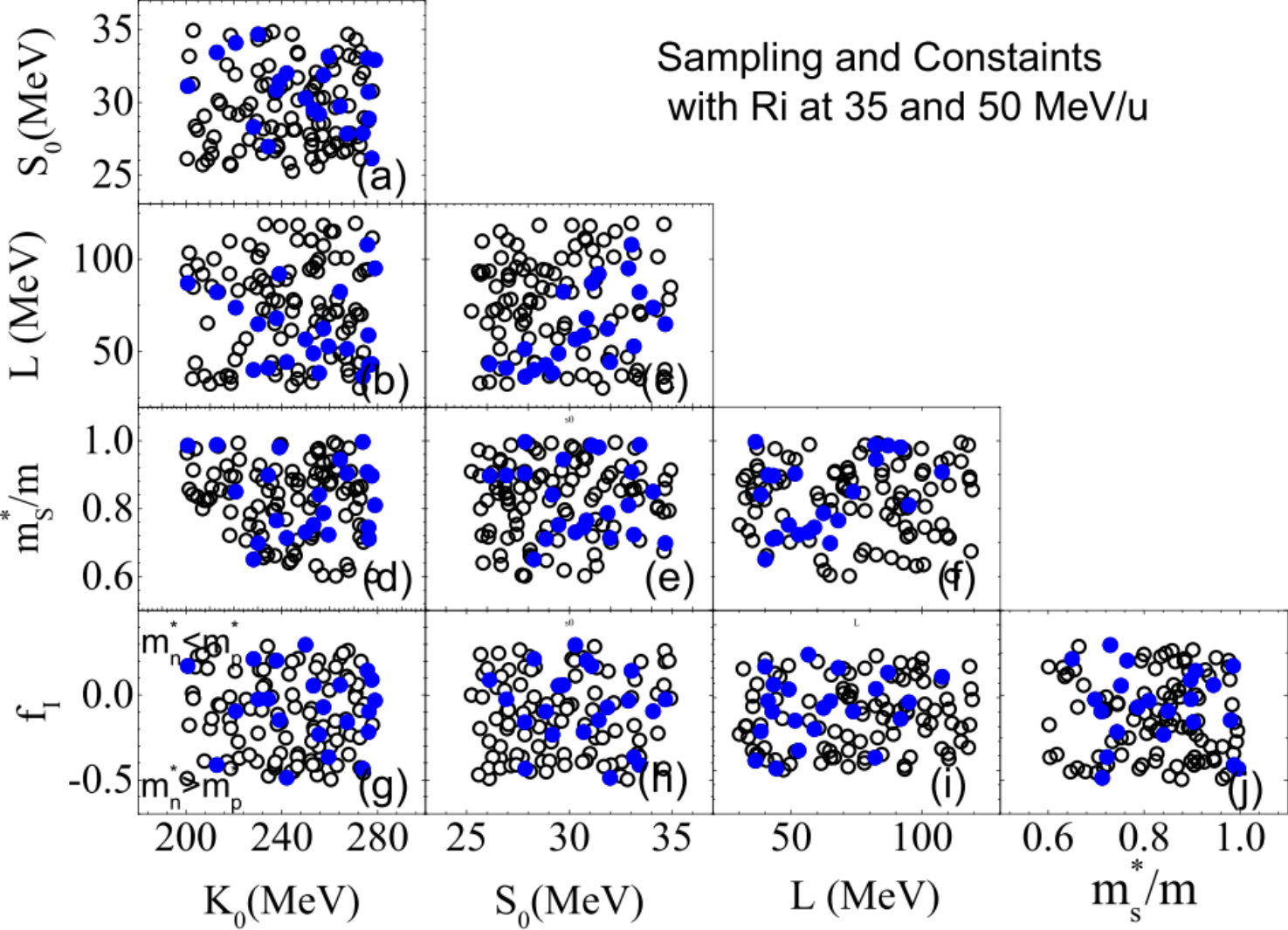}
\setlength{\abovecaptionskip}{0pt}
\caption{\label{samp-120}(Color online) Sampled points in 5D parameter space, blue solid points are the sets which can reproduce two isospin diffusion data.}
\setlength{\belowcaptionskip}{0pt}
\end{figure}

The values of the isospin transport ratio at projectile region reflect the isospin diffusion which depends on the stiffness of symmetry energy and the strength of the effective mass~\cite{LWChen05,Zhang15}. In this work, we investigate the isospin diffusion in five-dimensional (5D) parameter space, such as $K_0$, $S_0$, $L$, $m_s^*$, $f_I$. 
We sampled 120 points in 5D parameter space in the range which we listed in Table~\ref{tab:table1} under the condition that $\eta\ge 1.1$. $\eta\ge 1.1$ is used for guaranteeing the reasonable three-body force in the transport model calculations. The ranges of these nuclear matter parameters are chosen based on the $prior$ information of Skyrme parameters as shown in Fig. ~\ref{sky-prior}. As an example, the 120 sampled points are presented as open and solid circles in two-dimensional projection in Fig.~\ref{samp-120}. The points of parameter sets uniformly distribute in two-dimensional projection except for the plots of $K_0$ and $m^*_s/m$ due to the restriction of $\eta\ge1.1$. We perform the calculations for isospin transport diffusion at 35 MeV/u and 50 MeV/u at b=5-8 fm with the impact parameter smearing~\cite{Lili18} for $^{112,124}$Sn+$^{112,124}$Sn. 10,000 events are calculated for each point in the parameter space and simulations are stopped at 400fm/$c$. The calculations are performed on TianHe-1 (A), the National Supercomputer Center in Tianjin.

\begin{table}[htbp]
\caption{\label{tab:table1}%
Model parameter space used in the codes for the simulation of $^{112,124}$Sn+$^{112,124}$Sn reaction. 120 parameter sets are sampled in this space by using the Latin Hyper-cuber Sampling method.}
\begin{ruledtabular}
\begin{tabular}{lcc}
\textrm{Para. Name}& \textrm{Values}& \textrm{Description}\\

\colrule
$K_0$ (MeV) & [200,280] & Incompressibility \\
$S_0$ (MeV) & [25,35] & Symmetry energy coefficient \\
$L$ (MeV) & [30,120] & Slope of symmetry energy \\
 $m_s^*/m_0$ & [0.6,1.0] & Isoscalar effective mass \\
$f_I=(\frac{m_0}{m_s^*}-\frac{m_0}{m_s^*})$ & [-0.5,0.4] & $f_I=\frac{1}{2\delta}(\frac{m_0}{m_n^*}-\frac{m_0}{m_p^*})$ \\
\end{tabular}
\end{ruledtabular}
\end{table}

\begin{figure}[htbp]
\centering
\includegraphics[angle=0,scale=0.28]{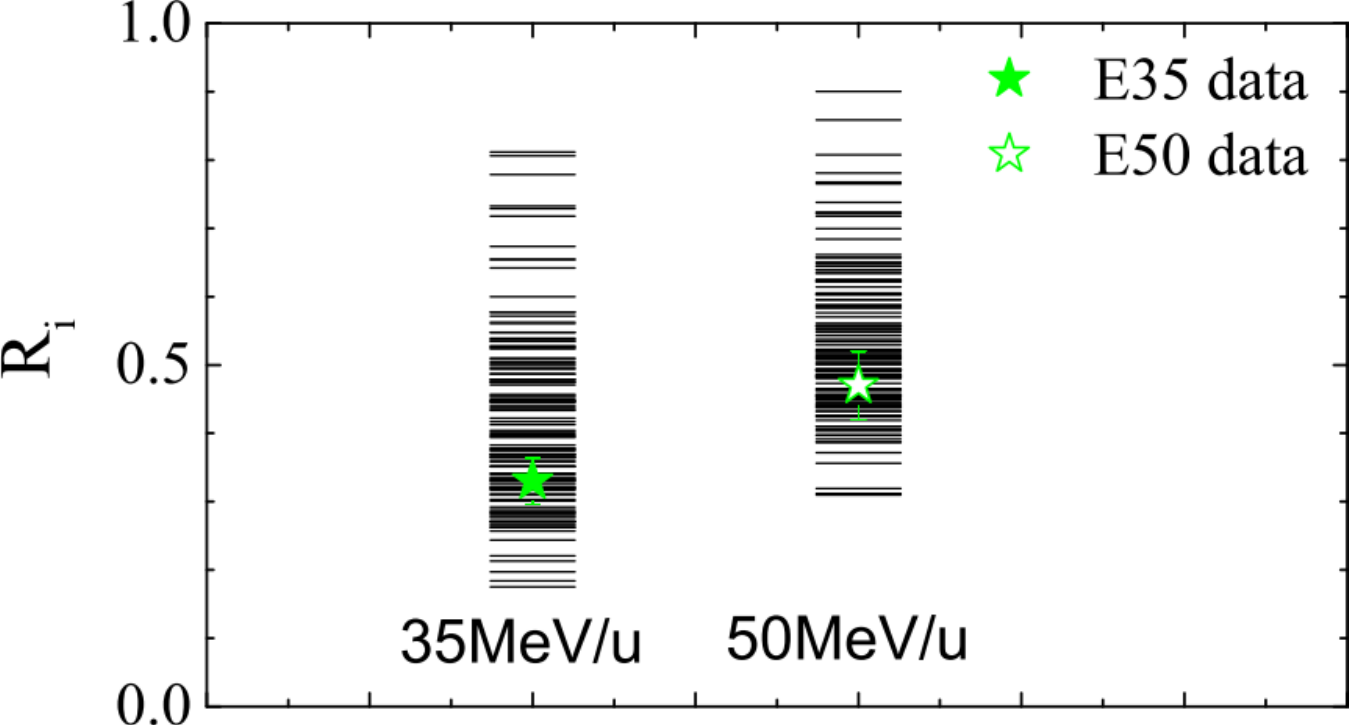}
\setlength{\abovecaptionskip}{0pt}
\caption{\label{ri-theexp}(Color online) Stars are the isospin diffusion data at 35 MeV/u and 50 MeV/u~\cite{Sun10, TXLiu07}, lines are the calculated isospin transport ratios with 120 parameter sets.}
\setlength{\belowcaptionskip}{0pt}
\end{figure}

In Fig.~\ref{ri-theexp}, the lines represent the calculated results of the isospin transport ratio $R_i$ with 120 parameter sets. Two stars are the experimental data~\cite{Sun10,Tsang04,TXLiu07} which is constructed from the isoscaling parameter $X=\alpha_{iso}$ at 50 MeV/u~\cite{Tsang04} and the ratio of $X=ln(Y(^7Li)/Y(^7Be))$~\cite{Sun10, TXLiu07} at the beam energy of 35 MeV/u, which was assumed and evidenced to linearly related to the isospin asymmetry of emitting source~\cite{TXLiu07}. And thus, one can compare the $R_i (\delta) $ to $R_i(\alpha)$ or $R_i(ln(Y(^7Li)/Y(^7Be))$. As shown in Fig.~\ref{ri-theexp}, the calculated results show a large spread around the experimental data. By comparing the calculations to the data, we find 22 parameter sets that can reproduce the isospin diffusion data within experimental errors. We highlight those points that can reproduce the experimental data within experimental errors with blue colors in Fig.~\ref{samp-120}. Generally, one can observe $L$ increases with $S_0$. The constrained points distribute in the bottom-right corner in the $S_0$-$L$ plot [panel (c)], and the large $L$ with small $S_0$ are ruled out.

\begin{figure}[htbp]
\centering
\includegraphics[angle=0,scale=0.4]{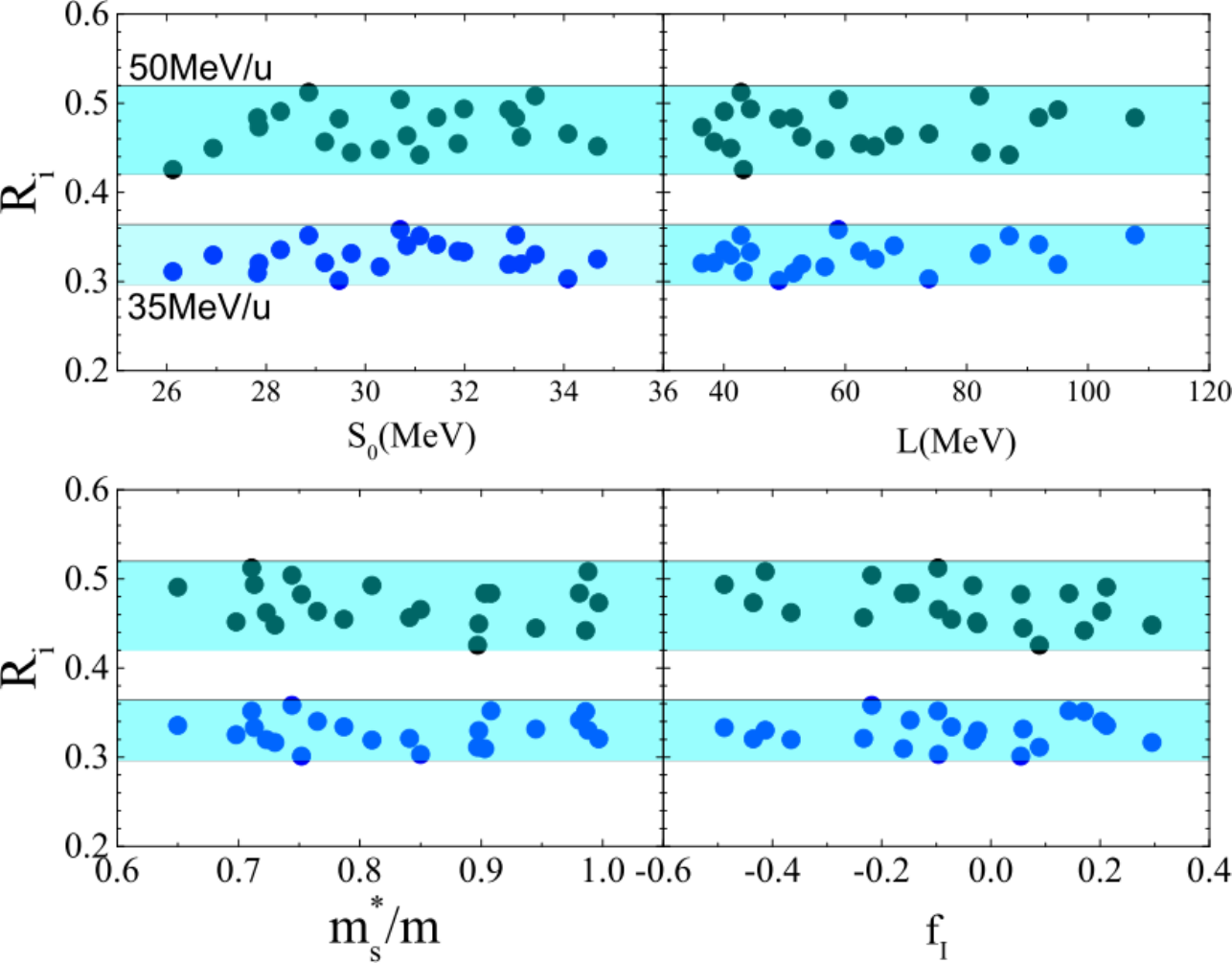}
\setlength{\abovecaptionskip}{0pt}
\caption{\label{ri-nmp}(Color online) $R_i$ as a function of $S_0$, $L$, $m_s^*/m$, and $f_I$, for the 22 points that can reproduce the isospin diffusion data.}
\setlength{\belowcaptionskip}{0pt}
\end{figure}

The results in panel (j) of Fig.~\ref{samp-120} show that isospin diffusion data is not sensitive to the effective mass and its splitting. In Fig.~\ref{ri-nmp}, we plot $R_i$ as a function of $S_0$, $L$, $m_s^*/m$, and $f_I$, and no obvious correlations between $R_i$ and $S_0$, $L$, $m_s^*/m$, and $f_I$ can be found. It is because the $R_i$ is not only correlated to $L$ but also correlated to $m_s^*/m$~\cite{Zhang15}, which broke the $R_i$ dependence of $L$ when we randomly chose the values of $K_0$, $S_0$, $L$, $m_s^*/m$, and $f_I$. If we fix the values of $K_0$, $S_0$, $m_s^*/m$, and $f_I$, the positive correlation between $R_i$ and $L$ can be found.

\subsection{Neutron skin and properties of neutron star}
Before calculating the neutron skin of $^{208}$Pb, i.e., $\Delta r_{np}$, with the RDV method, we first construct the effective standard Skyrme parameter sets, $t_0$, $t_1$, $t_2$, $t_3$, $x_0$, $x_1$, $x_2$, $x_3$, $\sigma$\, from the obtained nuclear matter parameters based on the Eq.~(\ref{eq:skyqmd}) and relations in Refs.~\cite{Zhang06,Zhang14}. In Table~\ref{tab:22skyrmes}, we present the extracted 22 standard Skryme parameter sets. The average values of nuclear matter parameters and its standard deviation from 22 sets are, $K_0$=250.54$\pm$22.87 MeV, $S_0$=30.62$\pm$2.39 MeV, $L$=62.31$\pm$21.01 MeV, $m_s^*/m$=0.83$\pm$0.11, and $f_I$=-0.072$\pm$0.22, which are consistent with previous constraints~\cite{Dutra12,BALi08,Mondal17,Margueron18}. Specially, the values of $f_I$=-0.072$\pm$0.22 mean $m_n^*>m_p^*$, it is consistent with the results from $ab initio$ calculations~\cite{Hassaneen04,ZYMa04,Dalen05,Holt16,Baldo17} and the analysis from the optical model analysis of nucleon-nucleus scattering~\cite{XHLi15,ZZhang16}. It seems contradictary with our analysis from the neutron to proton yield ratios~\cite{Pierre19}. But one should notice that both of our results on the effective mass splitting based on the heavy ion collision data have large uncertainties, the results suggest that we need a more critical examination in the future with new heavy ion collision observables.

\begin{figure}[htbp]
\centering
\includegraphics[angle=0,scale=0.3]{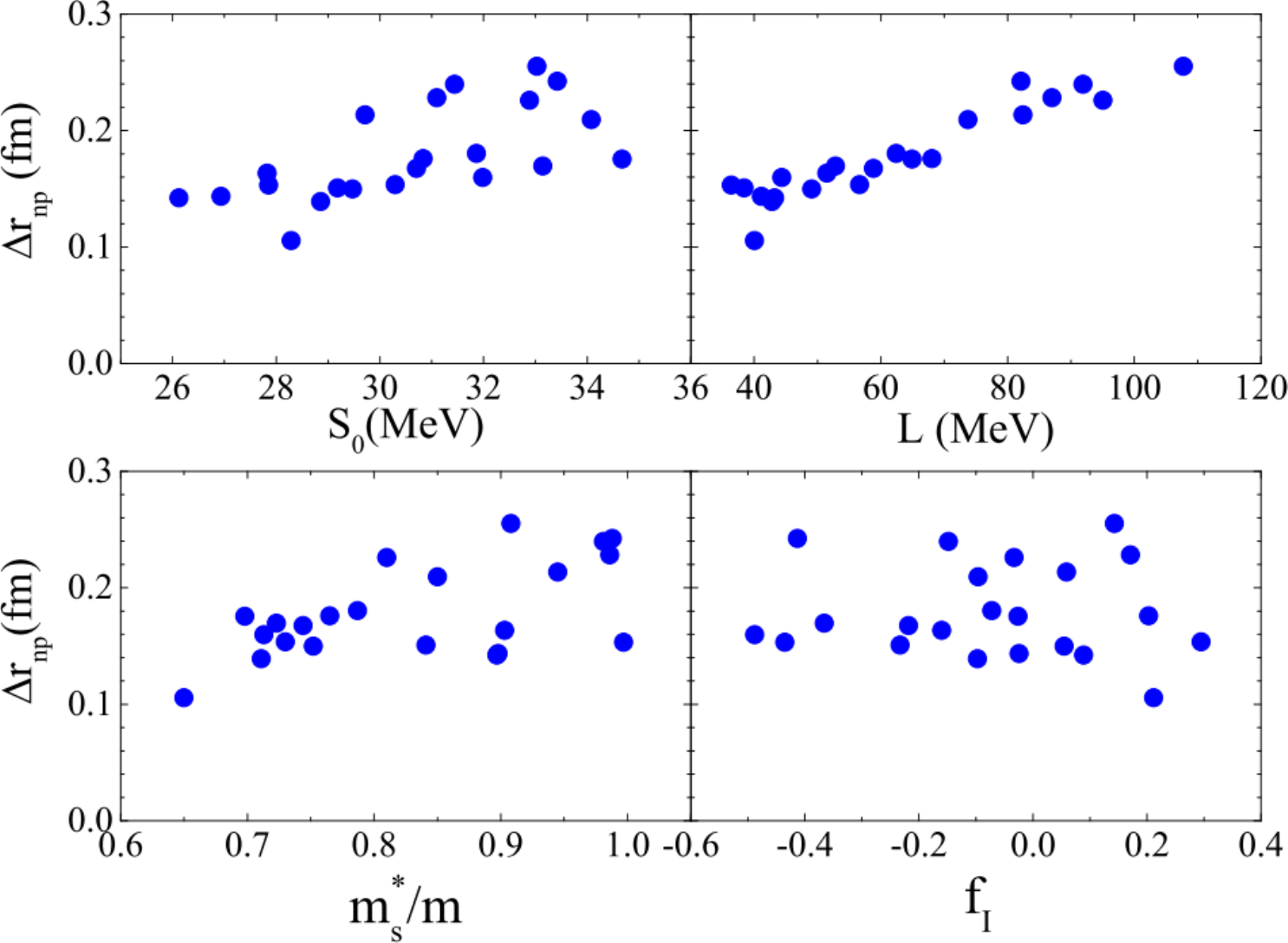}
\setlength{\abovecaptionskip}{0pt}
\caption{\label{nskin-nm}(Color online) $\Delta r_{np}$ obtained with 22 sets as a function of $S_0$, $L$, $m_s^*/m$, $f_I$.}
\setlength{\belowcaptionskip}{0pt}
\end{figure}

\begin{figure}[htbp]
\centering
\includegraphics[angle=0,scale=0.28]{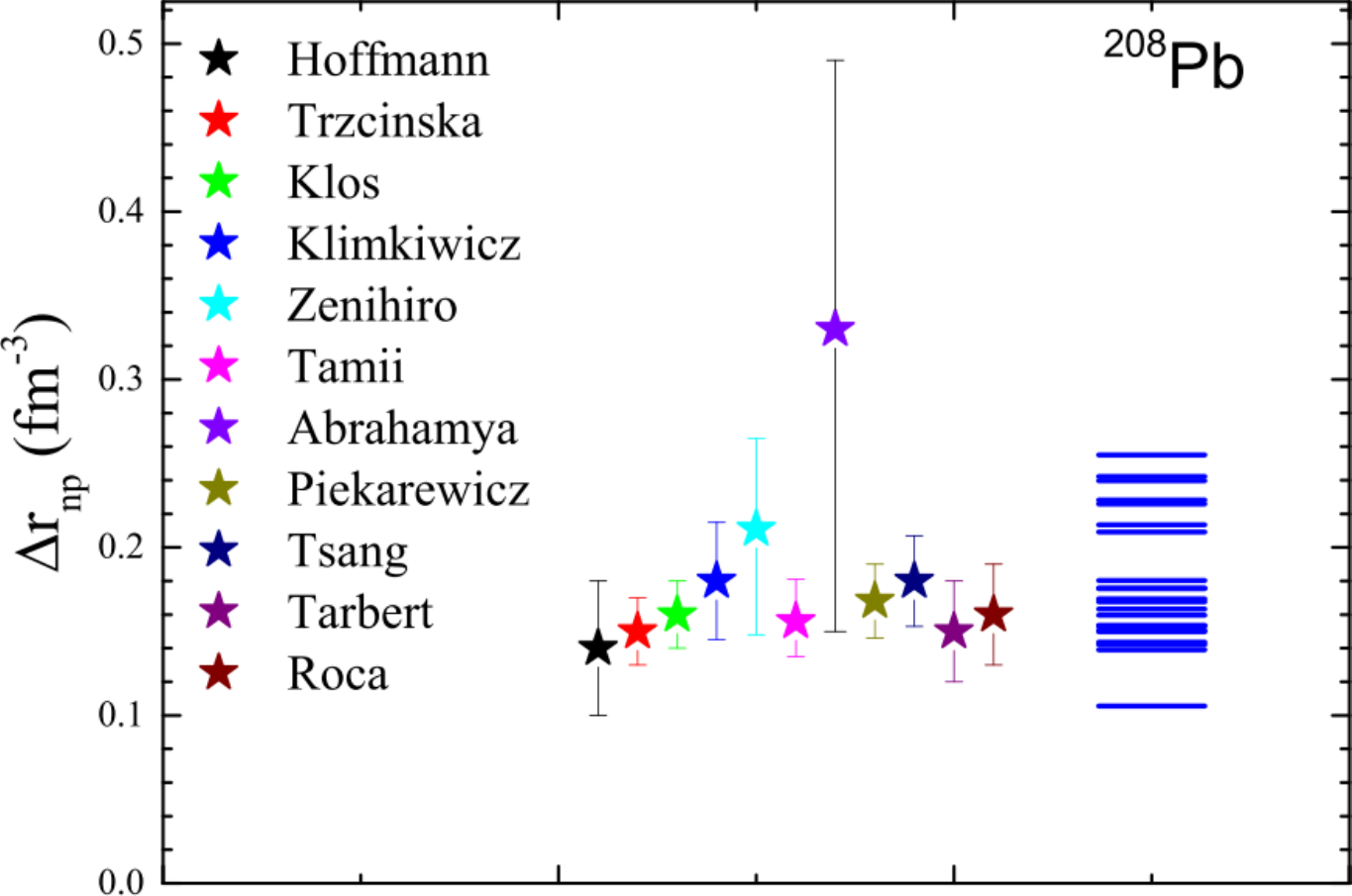}
\setlength{\abovecaptionskip}{0pt}
\caption{\label{nskin}(Color online) Symbols are the data of $\Delta r_{np}$ obtained from different groups, lines are the results of $\Delta r_{np}$ obtained with 22 parameter sets with the RDV method.}
\setlength{\belowcaptionskip}{0pt}
\end{figure}

In the calculations of neutron skin with the RDV method, $W_0=130$ MeVfm$^5$ is used. Figure~\ref{nskin-nm} presents the $\Delta r_{np}$ as a function of $S_0$, $L$, $m_s^*/m$, and $f_I$. The correlations between $\Delta r_{np}$ and $m_s^*/m$ ($f_I$) are very weak. The obvious correlation is between $\Delta r_{np}$ and $L$, which exists even we vary other nuclear matter parameters, such as $K_0$, $S_0$, $m_s^*/m$, and $f_I$. It is consistent with the results in Refs.~\cite{Centelles09,LWChen10,Mondal17,Roca13}. In Fig.~\ref{nskin}, the lines represent the neutron skin of $^{208}$Pb obtained by using 22 parameter sets. The stars are the neutron skin values extracted from the different groups~\cite{Hoff80, Trzc01, Klos07, Klim07, Zeni10, Tamii11, Abra12, Piek12, Tarb14, Roca15,Tsang12}. Within the large errors from PREX~\cite{Abra12}, all of our calculations fall into the data uncertainties. The 22 parameter sets also can give the prediction of neutron skin. We calculate the averaged values of the neutron skin of $^{208}$Pb and its value is $\Delta r_{np}=0.179\pm0.040$ fm, and it is consistent with the neutron skin values extracted from the experiments in Refs.~\cite{Hoff80, Trzc01, Klos07, Klim07, Zeni10, Tamii11, Abra12, Piek12, Tarb14, Roca15,Tsang12}. On another side, the strong and robust correlation between $\Delta r_{np}$ and $L$ as shown in Fig.~\ref{nskin-nm}, also suggest to us a precise measurement of neutron skin values could help us tightly constrain the symmetry energy at subsaturation density.

\begin{table*}[htbp]
\caption{\label{tab:22skyrmes} Extracted 22 nuclear matter parameter sets and the corresponding standard Skyrme parameters. $t_0$ in $MeVfm^3$, $t_1$ and $t_2$ in $MeVfm^5$, $t_3$ in $MeVfm^{3\sigma+3}$, $x_0$ to $x_3$ is dimensionless quantities. In the RDV calculations of this work, $W_0=130 MeVfm^{5}$ and $\rho_0=0.16fm^{-3}$.}
    \begin{tabular}{ccccccccccccccc}
    \hline
    \hline
No. & $K_0$ & $S_0$ & $L$ & $m_s^*/m$ & $f_I$ &$t_0$ & $t_1$  & $t_2$  & $t_3$ & $x_0$ & $x_1$ & $x_2$ & $x_3$ & $\sigma$ \\
\hline
1& 234.391 & 26.936 & 41.147 & 0.898 & -0.024 &-1890.80 &427.97 &-490.81 &	12571.72 &	0.10669 &	-0.19396 &	-0.7161 &	0.15416 &	0.29804 \\
2& 277.553 & 26.124 & 43.235 & 0.897 & 0.089 & -1374.17 &428.19 &-607.42 &	10814.29 &	0.04292 &	-0.26258 &	-0.81939 &	0.24329 &	0.51892 \\
3& 259.484 & 33.146 & 52.855 & 0.723 & -0.366 & -1569.42&	474.60&	3.93&	9415.46&	0.21035&	-0.03708&	-41.13867&	-0.02844&	0.37265 \\
4& 257.436 & 31.863 & 62.418 & 0.787 & -0.072 & -1572.00&	455.14&	-359.50&	10186.44&	0.10568&	-0.18487&	-0.69112&	0.07323&	0.38608 \\
5& 249.937 & 30.298 & 56.647 & 0.73 & 0.295 & -1714.97&	472.30&	-688.83&	10110.07&	0.34791&	-0.39789&	-1.01437&	0.97341&	0.31666 \\
6& 267.291 & 27.828 & 51.482 & 0.903 & -0.16 & -1452.20&	426.91&	-352.89&	10979.89&	-0.02416&	-0.11056&	-0.50064&	-0.25793&	0.46733 \\
7& 276.418 & 28.86 & 42.831 & 0.711 & -0.097 & -1395.03&	478.63&	-263.07&	8737.27&	0.20269&	-0.18678&	-0.68719&	0.48667&	0.47509 \\
8& 200.821 & 31.098 & 87.039 & 0.986 & 0.171 & -3048.33&	410.78&	-744.73&	19381.38&	-0.28089&	-0.3043&	-0.8462&	-0.35056&	0.16036 \\
9& 228.2 & 28.292 & 40.048 & 0.65 & 0.212 & -3312.92&	501.46&	-515.21&	17988.52&	1.00059&	-0.36089&	-1.06232&	1.48966&	0.10376 \\
10& 253.203 & 29.474 & 49.084 & 0.752 & 0.055 & -1644.99&	465.37&	-460.59&	10070.75&	0.24038&	-0.26259&	-0.86375&	0.55912&	0.34745 \\
11& 242.098 & 31.985 & 44.36 & 0.713 & -0.488 & -1914.52&	477.95&	140.60&	10865.66&	0.15117&	0.02588&	-2.31398&	-0.12133&	0.25704 \\
12& 239.014 & 31.441 & 91.905 & 0.981 & -0.148 & -1766.26&	411.68&	-411.04&	12629.01&	-0.43493&	-0.10372&	-0.52328&	-0.93988&	0.34248 \\
13& 230.13 & 34.676 & 64.931 & 0.698 & -0.026 & -2480.04&	483.17&	-323.15&	13757.39&	0.39189&	-0.22784&	-0.82337&	0.54526&	0.16807 \\
14& 220.763 & 34.081 & 73.762 & 0.85 & -0.096 & -2359.49&	438.85&	-383.47&	14591.08&	-0.02704&	-0.15899&	-0.63047&	-0.17633&	0.20869 \\
15& 237.836 & 30.837 & 68.072 & 0.765 & 0.203 & -1945.23&	461.46&	-625.89&	11613.88&	0.15946&	-0.34378&	-0.95171&	0.41995&	0.26249 \\
16& 276.165 & 30.705 & 58.846 & 0.744 & -0.218 & -1393.55&	467.84&	-169.89&	9181.01&	0.06398&	-0.11247&	-0.2356&	-0.13504&	0.48318 \\
17& 212.881 & 33.425 & 82.13 & 0.988 & -0.413 & -2406.93&	410.43&	-139.80&	15831.81&	-0.50854&	0.06498&	0.90667&	-1.02398&	0.21879 \\
18& 273.816 & 27.854 & 36.382 & 0.997 & -0.435 & -1396.68&	408.85&	-121.72&	11646.34&	0.0986&	0.08113&	1.25635&	-0.43832&	0.51157 \\
19& 278.918 & 32.888 & 95.046 & 0.81 & -0.033 & -1368.43&	448.90&	-418.69&	9938.92&	-0.21753&	-0.20303&	-0.73659&	-0.6341&	0.51343 \\
20& 255.597 & 29.184 & 38.419 & 0.841 & -0.233 & -1579.20&	441.03&	-234.77&	10767.52&	0.19335&	-0.08009&	-0.25897&	0.09028&	0.39256 \\
21& 275.783 & 33.03 & 107.768 & 0.908 & 0.143 & -1386.09&	425.85&	-670.47&	10922.75&	-0.33682&	-0.29417&	-0.85204&	-0.68126&	0.51127 \\
22& 264.335 & 29.718 & 82.428 & 0.945 & 0.059 & -1474.21&	418.40&	-605.68&	11375.98&	-0.25086&	-0.23842&	-0.78177&	-0.4952&	0.45917\\
    \hline
& $\bar{K_0}$ & $\bar S_0$ & $\bar L$ & $ \bar {m_s^*}/m$ & $\bar f_I $ &$ \bar t_0 $ & $\bar t_1$  & $\bar t_2$  & $\bar t_3 $ & $\bar x_0$ & $\bar x_1$ & $\bar x_2$ & $\bar x_3$ & $\bar \sigma$ \\
Average &250.54 & 30.62 & 62.31 & 0.83 & -0.072 & -1838.43&	447.08&	-383.78&	11971.69&	0.05613 &	-0.17691 &	-2.4674 &	-0.0112 &	0.3534 \\
 error &(22.87) & (2.39) & (21.01) & (0.11) & (0.22) & (553.99) &	(28.05) &	(223.12) & (2783.59) &	(0.3255) &	(0.133) &	(8.66) &	(0.608) &	(0.130) \\
    \hline
    \hline
    \end{tabular}
\end{table*}

\begin{figure}[htbp]
\centering
\includegraphics[angle=0,scale=0.32]{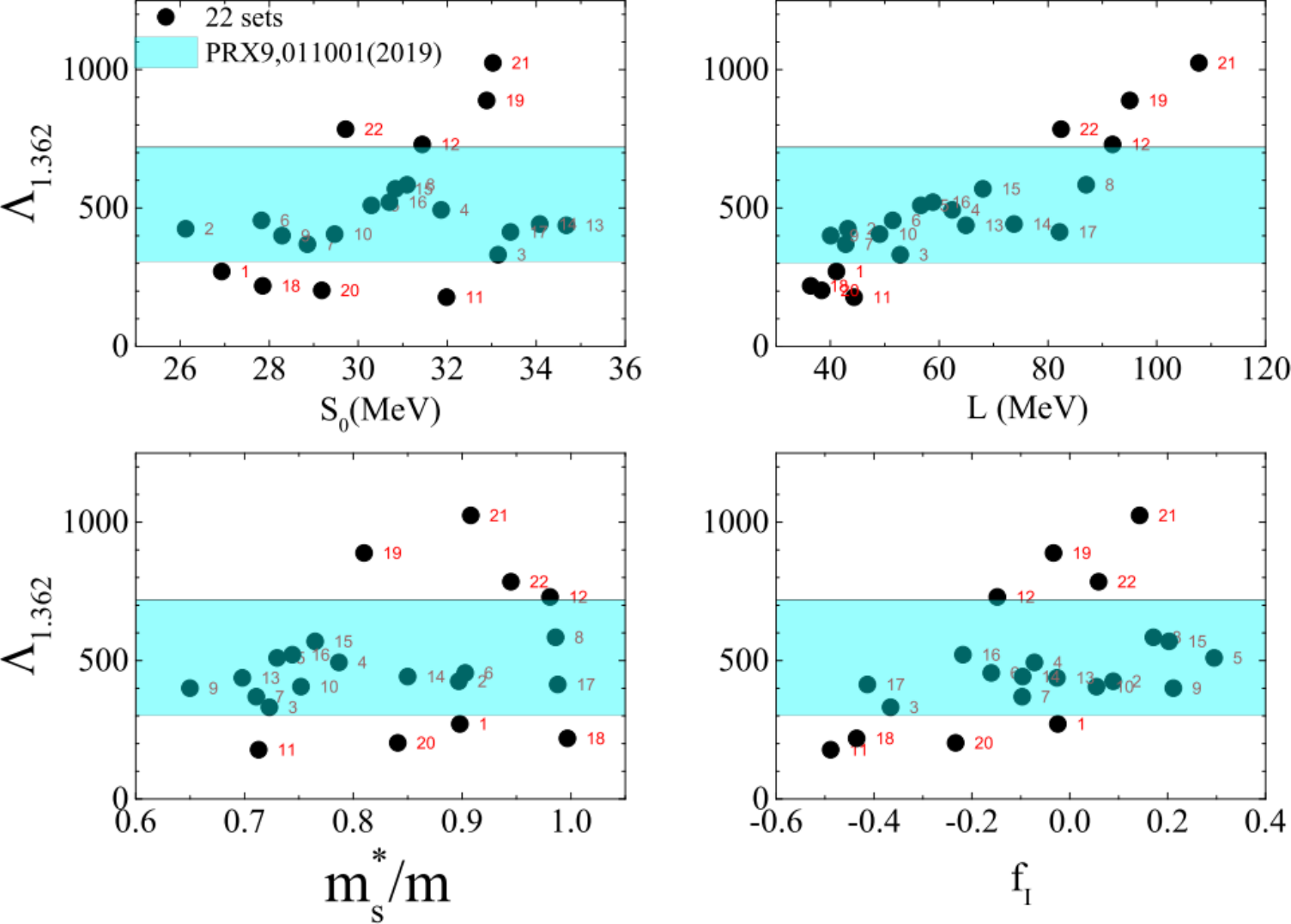}
\setlength{\abovecaptionskip}{0pt}
\caption{\label{lambda}(Color online) $\tilde\Lambda$ obtained with 22 parameter sets as a function of $S_0$, $L$, $m_s^*/m$, and $f_I$. The shadow region is the range of $\tilde\Lambda$ obtained with the binary neutron star merger event GW170817~\cite{Abbott17, Abbott19,Coughlin19,Carney18,De18,Chatziioannou18}.}
\setlength{\belowcaptionskip}{0pt}
\end{figure}

The structure of neutron stars is then obtained by solving the Tolman-Oppenheimer-Volkov equation, while the tidal deformability is estimated with $\Lambda = \frac{2 k_2}{3}\left( \frac{R}{G M} \right)^5$~\cite{Damour09, Hinderer10, Postnikov10}. Since the chirp mass for the binary neutron star merger event GW170817 is accurately measured with $\mathcal{M} = {(m_1 m_2)^{3/5}}{(m_1+m_2)^{-1/5}}=1.186\pm 0.001\ M_\odot$~\cite{Abbott19}, by assuming the mass ratio $m_2/m_1 = 1$, one obtains the dimensionless combined tidal deformability $\tilde{\Lambda}=\Lambda_1=\Lambda_2 \approx \Lambda_{1.362}$. The very recent constraint on $\tilde \Lambda$ is $302\leq \tilde{\Lambda} \leq 720$~\cite{Abbott17, Abbott19,Coughlin19,Carney18,De18,Chatziioannou18}, where an insignificant deviation from $\Lambda_{1.362}$ is expected for $\tilde{\Lambda}$. In Fig.~\ref{lambda}, we present the obtained dimensionless tidal deformability at $M=1.362\ M_\odot$ as a function of $S_0$, $L$, $m^*_s$ and $f_I$. The shadow region is the constrained $\tilde \Lambda$ values obtained in~\cite{Abbott17, Abbott19,Coughlin19,Carney18,De18,Chatziioannou18}. Our results show the $\tilde{\Lambda}$ is correlated with $L$, but weakly correlated to other nuclear matter parameters, such as $S_0$, $L$, $m^*_s$ and $f_I$.

\begin{figure}[htbp]
\centering
\includegraphics[angle=0,scale=0.32]{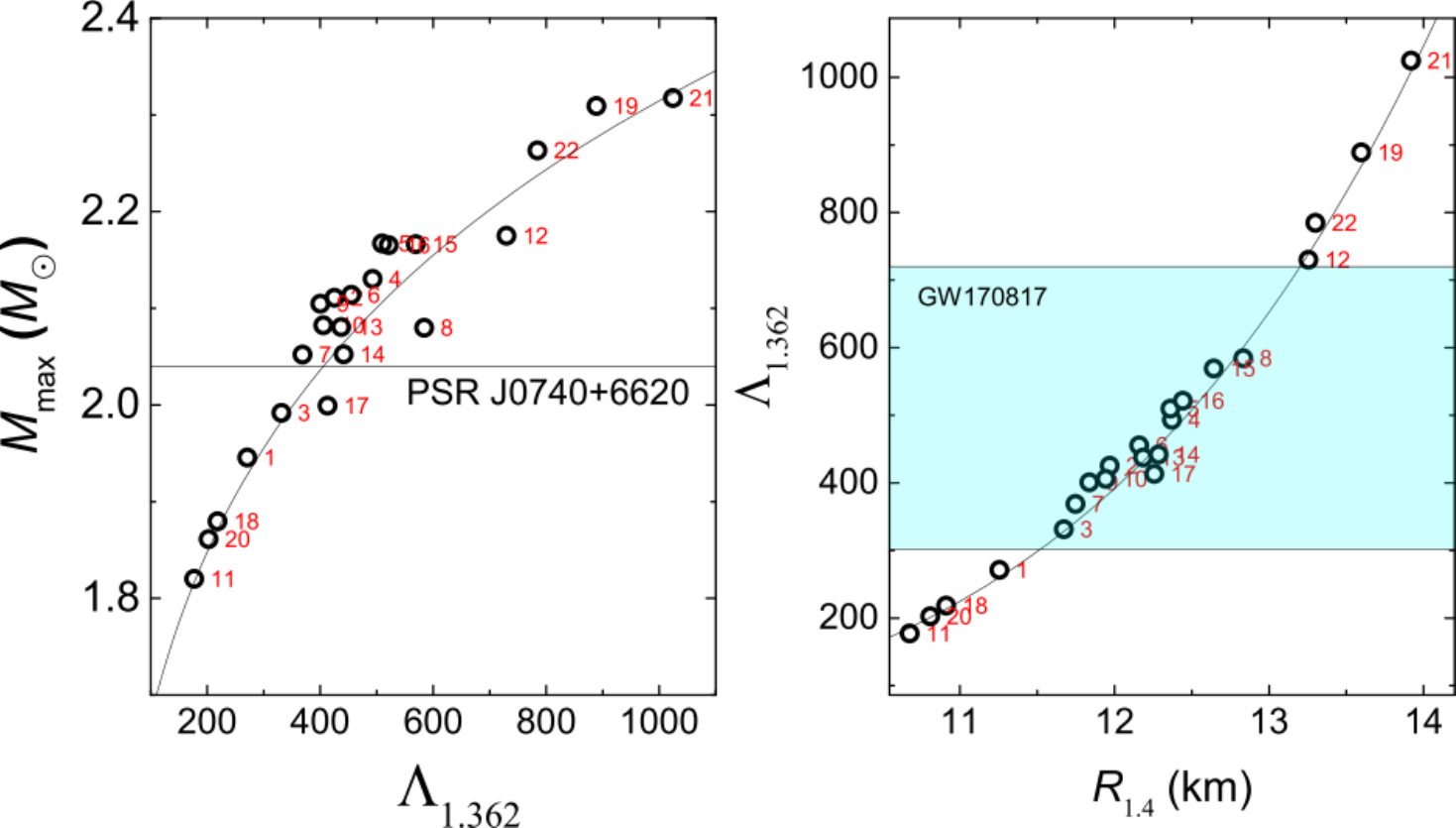}
\setlength{\abovecaptionskip}{0pt}
\caption{\label{lambda-r-max}(Color online)  Left panel is $M_{max}$ as a function of $\Lambda_{1.362}$, and right panel is $\Lambda_{1.362}$ as a function of $R_{1.4}$. The shadow region is the range of $\tilde\Lambda$ obtained with the binary neutron star merger event GW170817.}
\setlength{\belowcaptionskip}{0pt}
\end{figure}

As shown in Fig.~\ref{lambda-r-max}, the values of $\tilde{\Lambda}$ can be well reproduced by $\tilde{\Lambda}\approx 5.1\times 10^{-5} R_{1.4}^{6.38}$ with $R_{1.4}$ (in km) being the radius of a 1.4 solar mass neutron star. Within the constrained range of $\tilde \Lambda$, we have $11.5\leq R_{1.4} \leq 13.2$ km, which is in consistent with the recent radius measurements of PSR J0030+0451~\cite{Miller19,Riley19}.
A linear correlation between the maximum mass of neutron stars $M_\mathrm{max}$ (in $M_\odot$) and $\tilde{\Lambda}$ is observed with $M_\mathrm{max}\approx 0.88\tilde{\Lambda}^{0.14}$. Based on the observational mass of PSR J0740+6620 ($2.14{}_{-0.10}^{+0.09}\ M_\odot$)~\cite{Cromartie19}, a larger lower limit is then obtained with $\tilde{\Lambda}\gtrsim 370$, which is even larger than the center value reported in Ref.~\cite{Abbott19}. By calculating the $\tilde\Lambda$ and $M_{max}$, we finally find the parameter sets, Nos. 2, 4, 5, 6, 7, 8, 9, 10, 13, 14, 15, 16, in Table~\ref{tab:22skyrmes}, can satisfy the constraints of the neutron star.

\subsection{Symmetry energy and its related parameters}
\begin{figure}[htbp]
\centering
\includegraphics[angle=0,scale=0.35]{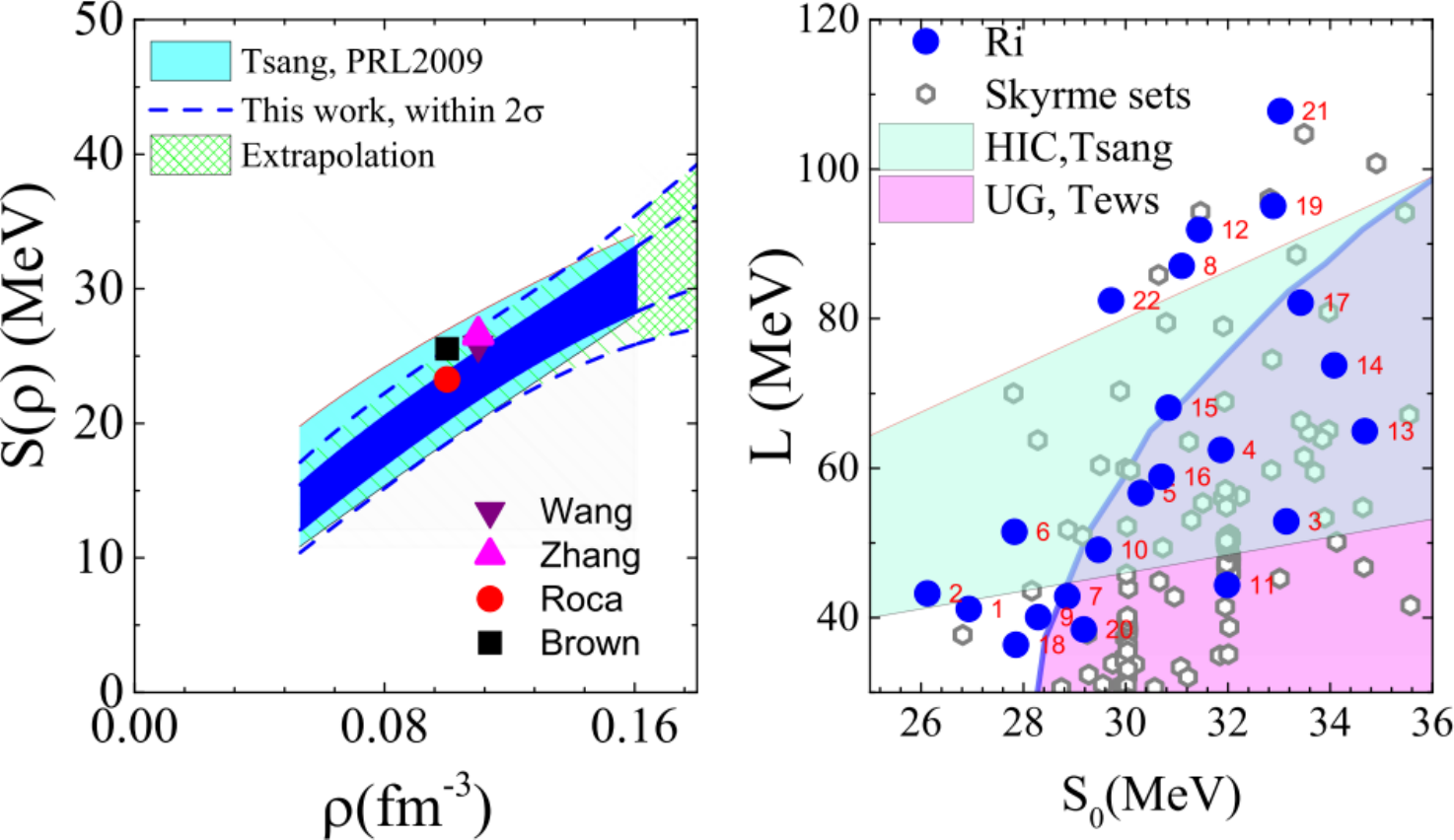} 
\setlength{\abovecaptionskip}{20pt}
\caption{\label{srho-ri}(Color online) Left panel is the constrained density dependence of symmetry energy in the range of 1/3-1.2 normal density. Right panel is the constraints of $S_0$ and $L$ values. The curved line in the right panel is the boundary of $S_0$ and $L$ obtained based on the unitary gas, and the pink region is allowed~\cite{Tews17}.}
\setlength{\belowcaptionskip}{20pt}
\end{figure}

Based on the extracted 22 Skyrme parameter sets, we can also obtain the corresponding symmetry energy according to Eq.~(\ref{srhoshf}). In the left panel of Fig.~\ref{srho-ri}, we present the obtained density dependence of the symmetry energy. The shadowed region with blue color represents for the $S(\rho)$ constrained from the two isospin diffusion data, i.e., $R_i$ at 35 MeV/u and 50 MeV/u, within 1$\sigma$. The region within the blue dashed lines is the constrained $S(\rho)$ within 2$\sigma$ uncertainties. The shadow region with cyan color is the constraint of the symmetry energy obtained in 2009 by analyzing the data of the isospin diffusion, isospin transport ratio, and double neutron to proton yield ratio at 50 MeV/u with ImQMD codes~\cite{Tsang09}, where the corresponding density dependence of the symmetry energy is
\begin{equation}
S(\rho)=\frac{1}{3}\frac{\hbar^2}{2m}(\frac{3\pi^2}{2})^{2/3}\rho^{2/3}+\frac{C_s}{2}(\frac{\rho}{\rho_0})^\gamma.
\end{equation}
Compared to the constraints of $S(\rho)$ by 2009 HIC data, the new analysis improves the constraints at the density below $\sim 0.13 fm^{-3}$ because we include isospin diffusion data at 35 MeV/u in this analysis. The uncertainties of the constraints of symmetry energy around normal density become larger than those in 2009, because the current analysis includes the uncertainties of $K_0$, $m^*_s$, and $f_I$. The symmetry energy obtained from the electric dipole polarizability in $^{208}$Pb~\cite{Roca13,ZZhang14}(red circle, up triangle), properties of doubly magic nuclei and masses of neutron-rich nuclei~\cite{Brown13} (black square) and Fermi-energy difference in finite nuclei~\cite{Wang13} (purple down triangle) are also presented in the left panel of Fig.~\ref{srho-ri}. The symmetry energy obtained in this work is also consistent with them~\cite{Brown13,ZZhang14,Wang13,Roca13,Roca15} within 2$\sigma$ uncertainties.

The consistence of the symmetry energy obtained from isospin diffusion data and the symmetry energy constraints from other nuclear structure studies~\cite{ZZhang14,Brown13,Wang13,Roca13} is because both of the isospin diffusion and nuclear structure studies reflect the information of symmetry energy at subsaturation density. It can be simply understood from the right panel of Fig.~\ref{srho-ri} by using the approach of sensitive density proposed by Lynch and Tsang~\cite{Lynch18}. 
The blue circle points in the right panel are the constraint by the isospin diffusion data at 35 and 50 MeV/u in this work. One can see there is a trend that $L$ increases with $S_0$, and the correlation between $S_0$ and $L$ is consistent with our previous work~\cite{Pierre19}.
By best linear fitting these points, the values of $\frac{\partial S_0}{\partial L}$ can be obtained, and we get $\frac{\partial S_0}{\partial L} = 0.061$ with standard error 0.022. Thus, the corresponding sensitive density is $\rho_s/\rho_0=0.685-0.946$ with $2\sigma$ of the $\frac{\partial S_0}{\partial L}$. The range of sensitive density is consistent with the dynamical prescription of the isospin diffusion process in peripheral heavy ion collisions, where the density in the neck region evolves from normal density to subnormal density until the neck breaks. This is also close to the corresponding average density region in the nuclear skin studies~\cite{Khan12,Brown13}.

The shaded cyan region in the right panel of Fig.~\ref{srho-ri} is the constrained $L$ and $S_0$ given by Tsang $et al$. in Ref.~\cite{Tsang09}, and the gray hexagon symbols are the parameters from the compilation of Dutra~\cite{Dutra12}.
We also present a boundary of symmetry energy parameters (thick curve) obtained in the unitary gas, and the pink region is the unitary gas is lower in energy than pure neutron matter~\cite{Tews17}. Under the constraints of this limit and previous constraints on $M_{max}$ and $\tilde\Lambda$, there are eight parameter sets that can describe the measured isospin diffusion, neutron skin, $M_{max}$, and $\tilde\Lambda$, which are Nos. 4, 5, 7, 10, 13, 14, 15, 16 in Table~\ref{tab:22skyrmes}. The average values of nuclear matter parameters and its standard deviations obtained with the eight parameter sets are, $K_0$=250.23$\pm$20.16 MeV, $S_0$=31.35$\pm$2.08 MeV, $L$=59.57$\pm$10.06 MeV, $m_s^*/m$=0.75$\pm$0.05, and $f_I$=0.005$\pm$0.170, which are consistent with our current knowledge of these parameters. One should notice the $f_I$ still have large uncertainties, and thus, the accurate knowledge of the sign and its magnitude of effective mass splitting still need to find more sensitive observables.


\section{Summary and outlook}

In summary, we study the symmetry energy and its related nuclear matter parameters by comparing the isospin diffusion data at 35 and 50 MeV/u to transport model calculations in five dimensional parameter space. We find the 22 parameter sets can well reproduce the data within the uncertainties of data. Our calculations show the positive correlation between $S_0$ and $L$ under the constraints from isospin diffusion data, and the $L$ values obtained from 22 parameter sets distribute from 30 to about 100 MeV. By using the 22 parameter sets, we calculate the neutron skin of $^{208}$Pb with the RDV method and obtain $\Delta r_{np}=0.179\pm 0.040$fm which is in the range of measured neutron skin. The strong and robust correlation between $\Delta r_{np}$ and $L$ is confirmed again, and it implies that a high precision data of neutron skin is needed and it will be very helpful for us to constrain the slope of symmetry energy.

Furthermore, the properties of neutron stars, such as the tidal deformability and maximum mass are also calculated and compared with the current constraints on $302\le \tilde\Lambda\le 720$ and $M_{max}> 2.14{}_{-0.10}^{+0.09}\ M_\odot$, we find there are only eight parameter sets can favor all the data of isospin diffusion, neutron skin, $\tilde \Lambda$, and $M_{max}$. The corresponding symmetry energy at 2$\rho_0$ is $S(2\rho_0)$=35-55 MeV which is consistent with the results in Refs.~\cite{NBZhang18,WJXie19,YZhou19APJ}. The average values of nuclear matter parameters and their standard deviations are calculated based on the 8 parameter sets, and we obtain $K_0$=250.23$\pm$20.16 MeV, $S_0$=31.35$\pm$2.08 MeV, $L$=59.57$\pm$10.06 MeV, $m_s^*/m$=0.75$\pm$0.05 and $f_I$=0.005$\pm$0.170. The estimated value of $f_I$ in this work is close to zero, which means the $m^*_n=m^*_p$, but we can not rule out $f_I>0$ or $f_I<0$ (i.e. $m^*_n<m^*_p$ or $m^*_n>m^*_p$) since the error of $f_I$ is huge. Thus, tightly constraining the isospin asymmetric nuclear equation of state and effective mass splitting may need more critical works in heavy ion collisions at different beam energies, and the measurement of neutron skin of nuclei, and mass-radius relations of neutron stars in future.


\begin{acknowledgements}
The authors thanks Prof. M. B. Tsang, Prof. H. St\"{o}cker, C. Y. Tsang, Prof. C. Mondal, and Prof. A. Li for the helpful discussions. This work was supported by the National Science Foundation of China Nos.11875323, 11875125, 11475262, 11705163, 11790323, 11790324, and 11790325, 11961141003, 11873040, the National Key R\&D Program of China under Grant No. 2018 YFA0404404, the Continuous Basic Scientific Research Project (No. WDJC-2019-13) and the funding of China Institute of Atomic Energy. The work was carried out at National Supercomputer Center in Tianjin, and the calculations were performed on TianHe-1 (A).
\end{acknowledgements}

\end{document}